# Coherent Spin Precession and Lifetime-Limited Spin Dephasing in CsPbBr$_3$ Perovskite Nanocrystals

Matthew J. Crane,[†,‡] Laura M. Jacoby,[†,‡] Theodore A. Cohen,[‡,§,∥] Yunping Huang,[§]

Christine K. Luscombe,[‡,§,∥] Daniel R. Gamelin[‡,∥,]*

[‡]*Department of Chemistry, University of Washington, Seattle, WA 98195-1700*
[∥]*Molecular Engineering & Sciences Institute, University of Washington, Seattle, WA 98195-1652*
[§]*Department of Materials Science & Engineering, University of Washington, Seattle, WA 98195-2120*

*Email: gamelin@chem.washington.edu

[†]*These authors contributed equally to this work.*

**Abstract.** Carrier spins in semiconductor nanocrystals are promising candidates for quantum information processing. Using a combination of time-resolved Faraday rotation and photoluminescence spectroscopies, we demonstrate optical spin polarization and coherent spin precession in colloidal CsPbBr$_3$ nanocrystals that persists up to room temperature. By suppressing the influence of inhomogeneous hyperfine fields with a small applied magnetic field, we demonstrate inhomogeneous hole transverse spin-dephasing times ($T_2^*$) that approach the nanocrystal photoluminescence lifetime, such that nearly all emitted photons derive from coherent hole spins. Thermally activated LO phonons drive additional spin dephasing at elevated temperatures, but coherent spin precession is still observed at room temperature. These data reveal several major distinctions between spins in nanocrystalline and bulk CsPbBr$_3$ and open the door for using metal-halide perovskite nanocrystals in spin-based quantum technologies.

Lead-halide perovskites are a promising class of materials with broad potential for optoelectronic applications stemming from their high photoluminescence quantum yields (PLQYs), chemically tunable bandgap, large absorption coefficients, and long carrier lifetimes.[1-6] They have also recently emerged as intriguing materials for future spintronic and quantum information applications due to their long spin lifetimes,[7, 8] strong spin-orbit coupling,[9] photoinduced spin polarization,[10, 11] Rashba effects,[12-14] and long optical coherence.[15, 16] Lead-halide perovskites offer the potential for generation, manipulation, and read-out of spins,[17] and have been proposed as promising candidates for applications such as spin field-effect transistors and single-photon emitters.[7, 16, 18, 19]

Recently, Utzat *et al*. demonstrated that $CsPbBr_3$ nanocrystals exhibit long optical coherence times, $T_{2,o}$, making them promising single-photon emitters.[16] A valuable figure of merit for single-photon emitters is the percentage of emitted photons that are optically coherent, which approaches 100% when $T_{2,o}$ reaches twice the PL lifetime, $\tau_{PL}$. In $CsPbBr_3$ nanocrystals, Rashba splitting creates a bright triplet state,[14] enabling fast radiative recombination at low temperatures and a $T_{2,o}/2\tau_{PL}$ ratio up to ~0.1, comparable to that of diamond color centers[16, 20, 21] and rivaling those of III-V epitaxial quantum dots.[22] Many other potential quantum-information capabilities, such as spin qubits, emerge in materials with long *spin* lifetimes and appropriate spectroscopic properties to allow all-optical spin generation, manipulation, and read-out.[23, 24] For example, some large-scale quantum-information schemes require quantum repeaters or spin-photon interfaces comprising single-photon emitters with long inhomogeneous transverse spin-dephasing times, $T_2^*$.[25-28] Thus, the combination of both spin and optical properties could make perovskite nanocrystals attractive for future quantum technologies.

Despite its importance, spin dephasing in lead-halide perovskites has only been investigated in bulk samples,[7, 8] and these do not show the long optical coherence times seen in nanocrystals. Additionally, studies of macroscopic samples don't address possibilities such as confinement-modified exchange interactions and exciton-binding energies,[29, 30] size-dependent Rashba effects,[31] modified phonon dispersion, and surface interactions,[32] all of which are critical determinants of spin dynamics in nanocrystals.[33] To date, spin dynamics in metal-halide perovskite nanocrystals have been investigated exclusively using zero-field circularly polarized transient-absorption spectroscopy.[9, 11] This technique measures longitudinal spin-flip dynamics, *i.e.*, zero-field $T_1$, whereas spin *dephasing* ($T_2$ or $T_2^*$) is the property of most interest for quantum applications. More detailed investigations of spins in lead-halide perovskite nanocrystals are thus needed.

Here, we use time-resolved Faraday rotation (TRFR) spectroscopy to probe spin-dephasing dynamics in CsPbBr$_3$ nanocrystals for the first time. We observe optical spin polarization and coherent precession of photogenerated holes, and identify two distinct dephasing regimes. At low temperatures, spin dephasing is driven by inhomogeneous hyperfine fields and can be suppressed by small magnetic fields, ultimately allowing detection of ensemble hole spin dephasing that is limited primarily by radiative carrier recombination. Exciton-phonon coupling drives spin dephasing at elevated temperatures, but spin coherence is still observed at room temperature in these nanocrystals, in contrast with bulk CsPbBr$_3$.[7, 8] The combination of short radiative lifetimes, long optical coherence times, and, now, long spin coherence times highlights the promise of CsPbBr$_3$ nanocrystals for spin-based quantum applications.

Figure 1a illustrates the one-color TRFR experiment used here.[17, 34-36] A resonant circularly polarized femtosecond pump pulse generates a polarized excited-state spin population that precesses in a transverse magnetic field and dephases. TRFR probes the ensemble spin projection at different delay times *via* Faraday rotation of a linearly polarized probe pulse. This experiment yields $T_2^*$, which provides a lower limit for the homogeneous spin coherence time, $T_2$.

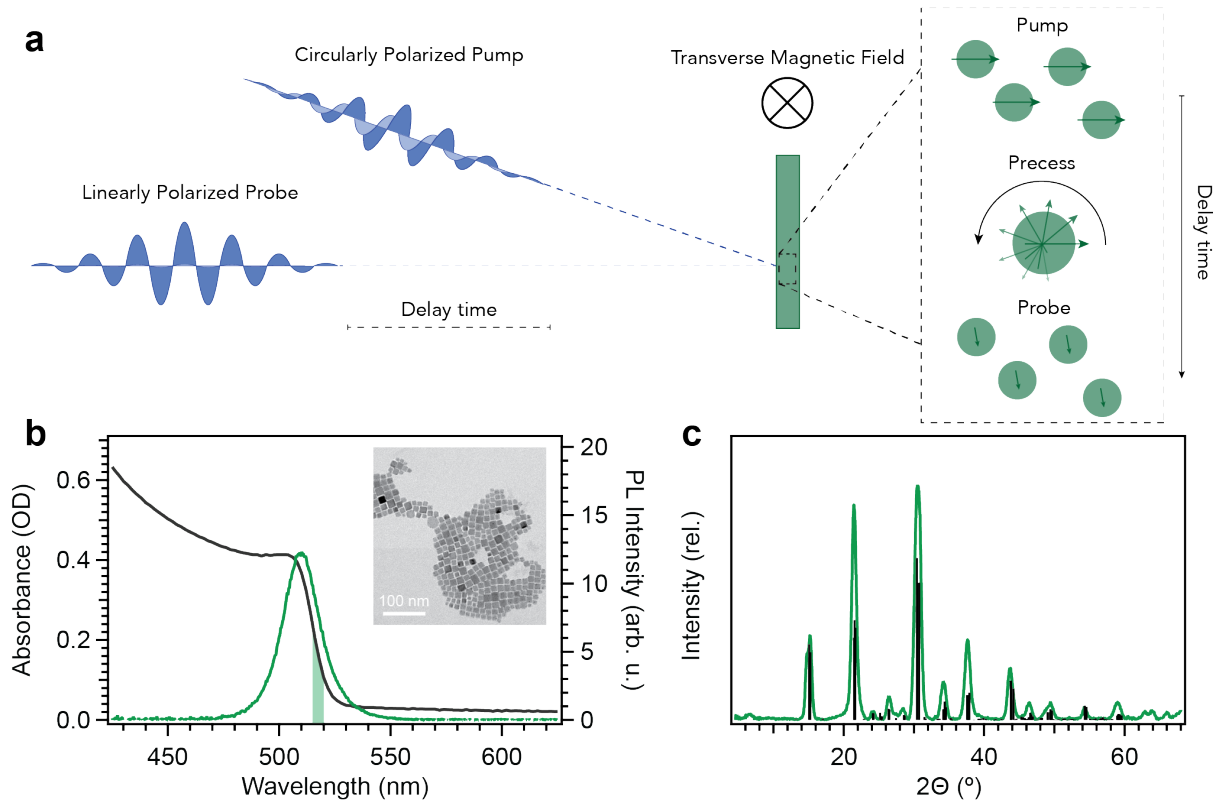

**Figure 1.** Time-resolved Faraday rotation (TRFR) experiment and $CsPbBr_3$ nanocrystal characterization. **(a)** A schematic summary of the TRFR experiment, illustrating optical generation of oriented spins in an ensemble of $CsPbBr_3$ nanocrystals using a circularly polarized femtosecond pump pulse resonant with band-edge absorption. The sample resides in a transverse magnetic field, causing the photogenerated spins to precess. A time-delayed linearly polarized probe pulse reads out the ensemble spin projection as a function of delay time, capturing spin-precession and dephasing dynamics. The panel on the right details the microscopic TRFR process for a single pump-probe cycle, highlighting the optical spin generation, spin precession, and read out *via* Faraday rotation of the linearly polarized probe. **(b)** Room-temperature absorption and PL spectra of $CsPbBr_3$ nanocrystals embedded in a thin polymer film (see SI). The shaded green region illustrates the pump and probe energies used in the TRFR measurements. The inset shows representative TEM images of the $CsPbBr_3$ nanocrystals without polymer. **(c)** Room-temperature XRD data for the $CsPbBr_3$ nanocrystals examined here. Black sticks illustrate reference reflections for orthorhombic $CsPbBr_3$. For variable-temperature

spectroscopic measurements described in the main text, these CsPbBr$_3$ nanocrystals were embedded in films of zwitterion-functionalized fluoropolymer (ZFP3)[37] on c-plane sapphire that showed negligible photon scattering (see SI for details). The TRFR measurement is extremely sensitive to the polarization of the probe pulse, and scattering scrambles this polarization information.

Figure 1b shows room-temperature absorption and PL spectra of representative CsPbBr$_3$ nanocrystals used in these TRFR experiments. From TEM (Fig. 1b, inset), these nanocrystals have edge lengths of 10 ± 2 nm, slightly larger than the Bohr exciton diameter of ~7 nm.[3] The sample shows bright room-temperature PL at 2.43 eV that is slightly Stokes shifted from the first excitonic absorption maximum (2.45 eV), with a room-temperature PLQY of ~30% under the TRFR conditions used here. Figure 1c plots X-ray diffraction (XRD) data collected for these nanocrystals, showing reflections consistent with the orthorhombic perovskite structure.

Figure 2a plots 4.5 K TRFR traces collected for these CsPbBr$_3$ nanocrystals at various transverse magnetic fields, $B_T$, from 0 to 0.60 T. At the highest field, the data show oscillations that decay within ~600 ps. At lower fields, the oscillation frequency decreases, and, at zero field, the trace shows only a non-oscillatory decay. At all non-zero fields, these oscillations can be associated with a single primary frequency, confirmed by the presence of single dominant peaks in the fast-Fourier transforms (FFT) of these traces (Fig. 2b, inset and Fig. S7, *vide infra*). These oscillations correspond to the Larmor precession frequencies ($\omega_L$) of the photogenerated spins and are related to the spin's *g* value by $\omega_L = g\mu_B B_T/\hbar$, where $\mu_B$ is the Bohr magneton. The TRFR amplitudes are proportional to the Faraday rotation angle, $\theta_F$, and reflect the sample's magnetization projected along the optical axis. The oscillations can be fit with the exponentially decaying oscillatory function given in eq 1.

$$\theta_F(t) = A\, e^{-t/T_2^*} \cos(\omega_L t + \phi) \qquad (1)$$

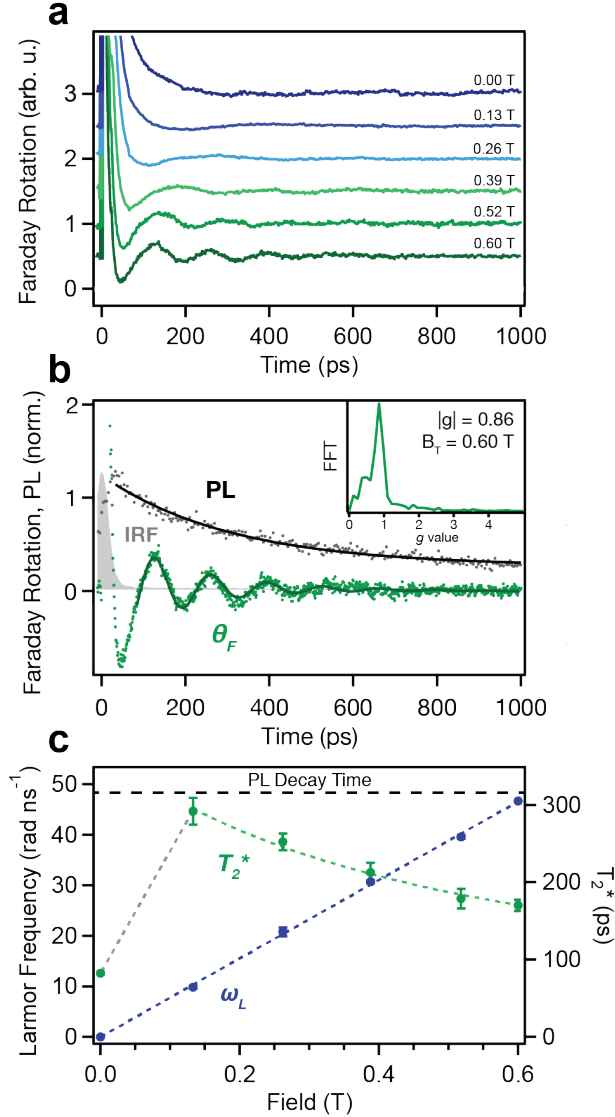

**Figure 2.** Spin dynamics in CsPbBr$_3$ nanocrystals measured at 4.5 K. **(a)** TRFR of CsPbBr$_3$ nanocrystals measured at different transverse magnetic field strengths. The oscillation frequency increases with increasing magnetic field. **(b)** Comparison of the TRFR trace measured at 0.60 T (green) and the PL intensity decay (black) measured under the same conditions ($\lambda_{ex}$ = 517 nm, T = 4.5 K). The gray curve corresponds to the instrument response function of the PL lifetime measurement. The solid curves show fits of the data. The TRFR trace is fit with a decaying sinusoidal function, yielding a decay time constant of $T_2^*$ = 170 ps. The PL decay is fit with a monoexponential function, yielding a decay time constant of $\tau_{PL}$ = 312 ± 11 ps. The inset plots the fast Fourier transform (FFT) of the TRFR trace and shows a single dominant resonance at $|g|$ = 0.86.

Detailed fits for each trace from panel (a) are provided as Supporting Information (Fig. S8). **(c)** Dependence of the Larmor frequency, $\omega_L$ (blue), and ensemble spin-dephasing time, $T_2^*$ (green), on the transverse magnetic-field strength, $B_T$. The Larmor frequency increases linearly with increasing magnetic field, and the slope (dashed blue line) gives $|g| = 0.88$. The initial increase in $T_2^*$ with applied field is attributed to suppression of hyperfine-induced local-field heterogeneity. The decrease in $T_2^*$ at larger applied fields is attributed to g-factor distribution, and the data from 0.13 to 0.60 T are fit (dashed green line) using eq 2, yielding $\Delta g = 0.07$. The gray dashed line between 0 and 0.16 T is a guide to the eye. For comparison, the black horizontal dashed line shows the PL decay time (312 ps) measured at 4.5 K under the same conditions.

Figure 2b shows a representative fit of the $B_T = 0.60$ T TRFR data with eq 1, which yields $\omega_L \sim 47$ rad ns$^{-1}$ and $T_2^* \sim 170 \pm 7$ ps. Figure 2b also compares the TRFR trace with the 4.5 K time-resolved PL of the same CsPbBr$_3$ nanocrystals measured under identical conditions. The PL decays monoexponentially with a lifetime of $\tau_{PL} = 312 \pm 11$ ps. Figure 2c plots the magnetic-field dependence of $\omega_L$ and $T_2^*$ obtained by fitting all of the TRFR traces in Fig. 2a (see Fig. S8 for fits). The magnitude of $\omega_L$ increases linearly with increasing $B_T$. Fitting this trend yields $|g| = 0.88$, which matches the g value obtained through FFT of these data ($|g| = 0.86$, Fig. 2b, inset, and Fig. S7). Based on first-principles models of metal-halide perovskite magneto-optics,[38, 39] we assign the measured species as precessing holes. The g value from Fig. 2c agrees well with the hole g value reported for bulk CsPbBr$_3$ ($|g| = 0.76$),[8] supporting this assignment. It also falls within the range of reported CsPbBr$_3$ nanocrystal hole g values (Table S2).[39, 40] FFT traces additionally show a weak, broad shoulder at $|g| \sim 2$ (see Fig. S9) that coincides with the electron g value of ~2 reported for bulk and nanocrystalline CsPbBr$_3$.[8, 39, 40] Although distinct, this signal is too weak to be thoroughly analyzed with confidence. The lack of a long-lived TRFR signal attributable to photogenerated electrons likely reflects rapid dephasing. Although rapid electron trapping[41] could conceivably be responsible, the short PL decay times combined with the near-unity PLQYs (*vide infra*) under these conditions suggest that electron spin dynamics in these nanocrystals are not dominated by trapping. We note that trapped carriers do not generate large band-edge TRFR signals.

It can be challenging to distinguish between excitons and trions because of their similar decay times at low temperatures.[40] Positive trions would show no hole precession and are therefore incompatible with the data. If negative trions dominated the TRFR signals, then only unpaired holes would be observed, but instead we also observe a distinct electron TRFR signal ($g_e$ ~ 1.8, Fig. S9). These considerations, combined with the nearly 100 % PLQY at this temperature (*vide infra*), thus strongly suggest that the holes probed by TRFR here are associated with excitons. This conclusion is consistent with the very low pump and probe pulse fluences used here (see SI) compared to those reportedly required for trion formation in $CsPbBr_3$ nanocrystals.[42]

At zero field, $T_2^*$ is ~80 ps, but application of a small magnetic field (0.13 T) increases $T_2^*$ to 292 ± 12 ps. Notably, this latter value is essentially indistinguishable from the PL decay time of 312 ± 11 ps measured under the same conditions (Fig. 2b). Above 0.13 T, $T_2^*$ decreases monotonically with increasing $B_T$, as commonly observed in the TRFR of other semiconductors.[36] The decrease in $T_2^*$ with increasing $B_T$ is attributable to a distribution in $g$ values ($\Delta g$), for example from a distribution in nanocrystal sizes or defect structures, and can be evaluated by fitting the higher-field $T_2^*$ data with eq 2.[36]

$$1/T_2^* = 1/T_2 + 1/T_2^{inh} \approx 1/T_2 + \Delta g \mu_B B_T/\hbar \qquad (2)$$

Here, $1/T_2^{inh}$ represents the sum of all inhomogeneous contributions to the spin-dephasing rate. This analysis yields $\Delta g$ = 0.07 (~8 %). The observation that $\Delta g$ here is similar to that in bulk lead-halide perovskites[7, 8] suggests that *g*-value heterogeneity stemming from the presence of different non-cubic crystal structures and nanocrystal orientations is not the dominant contributor to $T_2^*$. The observation that $T_2^*$ approaches but does not exceed the excitonic PL decay time further supports the conclusion that the holes probed by TRFR here are associated with excitons.

Short radiative lifetimes at low temperature are a distinctive characteristic of $CsPbBr_3$ and related lead-halide perovskite nanocrystals, whose lowest excitonic excited states are optically bright. This unique electronic structure enables the high $T_{2,o}/2\tau_{PL}$ ratios found for $CsPbBr_3$ nanocrystals.[14, 16] Significantly, Fig. 2 shows that $T_2^*$ is similar to $\tau_{PL}$, indicating that photogenerated hole spins retain their transverse coherence throughout a large portion of the excited-state lifetime. Consequently, a large fraction of the luminescence involves coherent polarized hole spins. Moreover, this fraction can be tuned using a magnetic field. For example, $T_2^*/\tau_{PL}$ at zero field indicates ~26% of emitted photons involve coherent hole spins, but increasing $B_T$ to 0.13 T raises this value to ~95% (see Fig. S4). The near equivalence of $T_2^*$ and $\tau_{PL}$ at 0.13 T indicates that even inhomogeneous hole transverse spin-dephasing times in $CsPbBr_3$ nanocrystals are primarily limited by recombination, and hence that the hole $T_2$ in, *e.g.*, $CsPbBr_3$ nanocrystal single-photon emitters, is entirely population limited.

To investigate the origins of hole spin dephasing in $CsPbBr_3$ nanocrystals, we explored the PL temperature dependence. Figure 3a shows a color map of nanocrystal PL intensities measured from 4.5 to 300 K. As the temperature is increased from 4.5 K, the near-band-edge PL peak broadens and shows an anti-Varshni blueshift characteristic of $CsPbBr_3$,[43] while decreasing in intensity. Figure 3b plots the integrated PL intensity ($I(T)$) and PL full-width-at-half-maximum (FWHM, $\Gamma(T)$) *vs* temperature. $I(T)$ is constant between 4.5 and ~100 K but decreases rapidly above 100 K. These data, in conjunction with the PLQY of ~30% at room temperature under these conditions, suggest that the PLQY between 4.5 and 100 K is close to 100%, as observed previously.[16, 44] We interpret the temperature dependence in Fig. 3b as thermally assisted exciton dissociation. In this scenario, the exciton-binding energy ($E_B$) can be estimated by fitting $I(T)$ with eq 3, where $I_0$ is the PL intensity at 0 K, $A$ is a prefactor, and $k_B$ is the Boltzmann constant.

$$I(T) = \frac{I_0}{1 + A\, e^{-E_B/k_B T}} \qquad (3)$$

This analysis yields $E_B = 64 \pm 2$ meV, similar to previous reports, as well as $A = 85$ and $I_0 = 3.05 \cdot 10^5$ counts.[29, 30] In addition, $\Gamma(T)$ can be analyzed by fitting the data in Fig. 3b with an independent Boson model described using eq 4, where $\Gamma_0$, $\sigma$, and $\Gamma_{op}$ reflect inhomogeneous PL broadening, an acoustic phonon contribution, and exciton-optical phonon coupling with energy $\hbar\omega_{op}$, respectively.[45]

$$\Gamma(T) = \Gamma_0 + \sigma T + \frac{\Gamma_{op}}{e^{-\hbar\omega_{op}/k_B T} - 1} \qquad (4)$$

This analysis reveals exciton coupling to longitudinal optical (LO) phonons with energy $\hbar\omega_{op} = 36 \pm 8$ meV, as well as $\Gamma_0 = 32 \pm 1$ meV, $\sigma = 60 \pm 20$ µeV, and $\Gamma_{op} = 80 \pm 12$ meV, consistent with other perovskite nanocrystal reports (Table S4).[30, 46, 47]

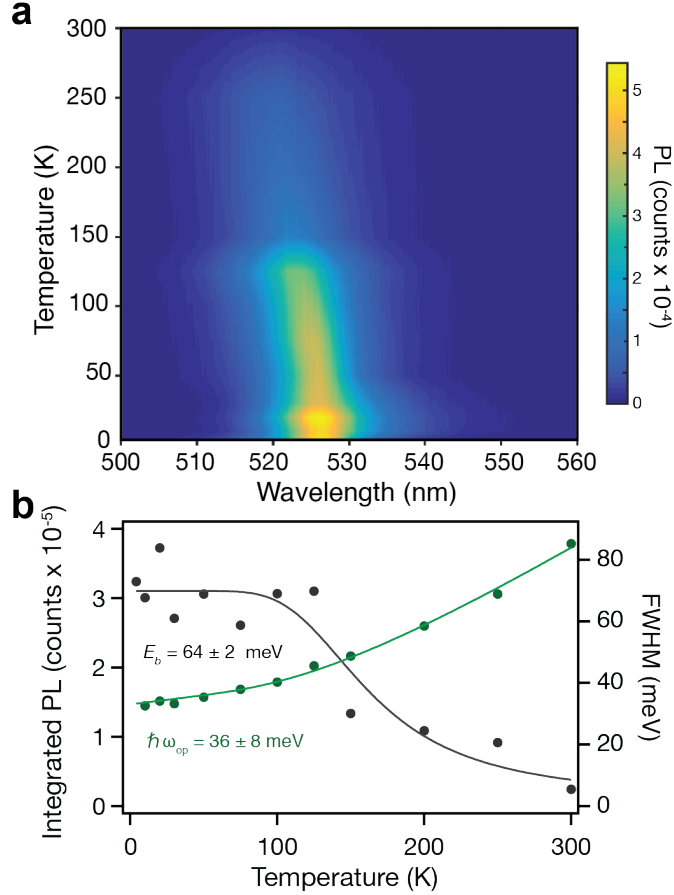

**Figure 3.** (a) Variable-temperature PL intensities for the CsPbBr$_3$/ZFP3 film used to collect TRFR data, measured using 375 nm excitation. (b) Integrated exciton PL intensity ($I(T)$, black) and FWHM ($\Gamma(T)$, green) plotted *vs* temperature. The solid curves show fits of the data using eq 3 and eq 4. Fitting the variable-temperature PL intensities yields an exciton binding energy of $E_B = 64 \pm 2$ meV. Fitting the exciton FWHM data yields an optical phonon energy of $\hbar\omega_{op} = 36 \pm 8$ meV.

Figure 4 plots $T_2^*$ *vs* temperature for data measured with and without a transverse magnetic field of 0.60 T. Three distinct regimes are identified in the temperature dependence of $T_2^*$. Below 50 K, $T_2^*$ is much larger in the transverse field than at zero-field. For example, at 4.5 K, $T_2^*$ is ~170 ps when $B_T = 0.60$ T but only ~80 ps when $B_T = 0$ T (see Fig. S6). Whereas the zero-field $T_2^*$ is independent of temperature in this regime, $T_2^*$ measured at 0.60 T decreases rapidly with increasing temperature until the two data sets converge at ~50 K. Between 50 and 100 K, $T_2^*$ is independent of both temperature and applied field. Above ~100 K, both data sets show $T_2^*$ decreasing with increasing temperature, dropping to ~16 ps at room temperature.

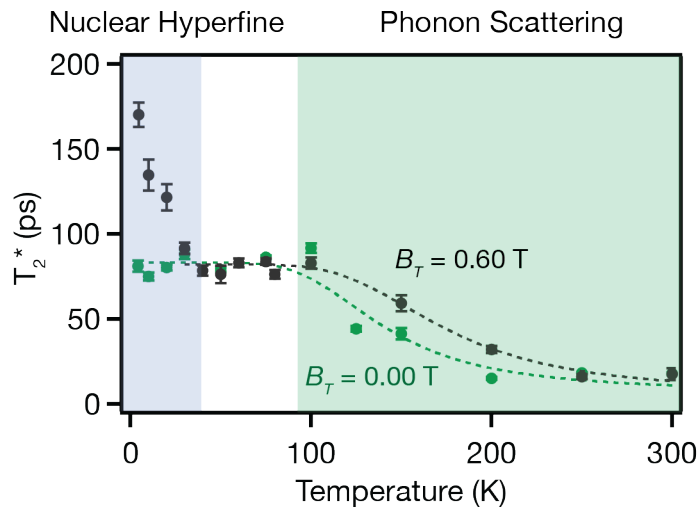

**Figure 4.** Ensemble inhomogeneous transverse spin-dephasing times ($T_2^*$) measured for CsPbBr$_3$ nanocrystals with and without an applied magnetic field, plotted as a function of sample temperature. The blue area highlights the regime where other dephasing mechanisms are frozen out and the applied magnetic field boosts $T_2^*$ by polarizing the nuclear spin bath. The green area emphasizes the regime where thermally activated phonon scattering dominates spin dephasing. The dashed green curve shows a fit to eq 5 of the entire zero-field $T_2^*$ data set. The dashed black curve shows a fit to eq 5 of the $T_2^*$ data collected at 0.60 T between 40 and 300 K. See Fig. S5, for time traces.

These observations suggest two dominant thermally activated spin-dephasing mechanisms in this material: a field-dependent mechanism active at low temperatures and a field-independent mechanism active at higher temperatures. We attribute the large increase in $T_2^*$ upon application of the transverse magnetic field below 50 K to field-induced suppression of local magnetic-field inhomogeneities due to nuclear hyperfine interactions that accelerate dephasing.[20, 33, 48-50] The highly ionic nature of the metal-halide perovskite lattice and the large Pb(6s) contribution to the carrier wavefunction at the valence-band edge favor strong Fermi-contact hyperfine coupling (Table S2).[50-52] Accumulation of Pb$^{2+}$ nuclear spin polarization may further exacerbate the effect of inhomogeneous nuclear magnetic fields.[8] Applied magnetic fields can also reduce dephasing caused by spontaneous electromagnetic fields generated by the Rashba effect, as observed in III-V nanomaterials.[53, 54]

The higher-temperature (100 - 300 K) dephasing is attributed to coupling between photogenerated spins and LO phonons. For example, at elevated temperatures and small $B_T$, elastic scattering of phonons can modulate the hyperfine field, causing decoherence.[55, 56] Similarly, the Elliott–Yafet mechanism[9, 57] involves spin-flip processes driven by phonon coupling, and the Dyakonov–Perel mechanism[58] involves inhomogeneous magnetic fields due to noncentrosymmetric local distortions, potentially driven by optical phonons in this case.[12, 59] The effect of phonon-modulated hyperfine fields on dephasing can be described by eq 5, where the phonon-induced decoherence term, $\Lambda$, is a function of the nuclear spin values, their concentration, the volume of the nanocrystal, and the hyperfine coupling strength. $T_2^*(0)$ represents the low-temperature $T_2^*$ at zero magnetic field, and $F(x) = (1 - \tanh^2(x))\tanh(x)$.

$$T_2^*(T) = \left(1/T_2^*(0) + \Lambda F\left(\frac{\hbar\omega}{k_B T}\right)\right)^{-1} \qquad (5)$$

This fitting yields $\hbar\omega = 29 \pm 5$ meV ($\Lambda = 0.27$ ps$^{-1}$) for the zero-field data and $\hbar\omega = 36 \pm 5$ meV ($\Lambda = 0.32$ ps$^{-1}$) for the 0.60 T data, which are within experimental uncertainty of one another. Both are similar to the LO phonon frequency that dominates exciton-phonon coupling ($\hbar\omega = 36 \pm 8$ meV, Fig. 3b). Exciton coupling with acoustic phonons is weak in CsPbBr$_3$ nanocrystals at low temperature,[43] which could explain its limited influence on spin dephasing in this temperature range. Thermally activated spin dephasing in other lead-halide perovskites has been described as Arrhenius-like and attributed to LO phonons with a temperature dependence similar to eq 5.[7-9]

Although the temperature dependence of $T_2^*$ in these CsPbBr$_3$ nanocrystals at zero field is qualitatively similar to that measured for bulk metal-halide perovskites, including CsPbBr$_3$ single crystals,[7, 8] the data here reveal several important contrasts between spins in nanocrystalline and bulk CsPbBr$_3$. Most notably, the photogenerated spins monitored by TRFR in bulk CsPbBr$_3$ reportedly[8, 32] accumulate over time, likely because of deep carrier trapping, such that they cannot be directly associated with the emissive excited state of the material. This conclusion is supported by the very low PLQYs and by $T_2^*$ values that exceed the PL decay time in these bulk samples.[29] Bulk samples also showed sizable variations in hole $T_2^*$ depending on excitation position, with values ranging from 0.7 to 1.9 ns.[8] In contrast, because of the combination of short radiative lifetimes and high PLQYs (~100% at 4.5 K) in the nanocrystals, the spins of photogenerated holes in CsPbBr$_3$ nanocrystals *are* associated with the luminescent excited state and appear to remain coherent for the entire lifetime of this excited state when under a small magnetic field, *i.e.*, until they recombine radiatively. Consistent with these observations, there is also essentially no variation in $T_2^*$ at different excitation positions within a given nanocrystal film or between nanocrystal samples (see Table S3, Fig. S3).

A second notable contrast is that $T_2^*$ shows a strong field dependence at low temperatures in the CsPbBr$_3$ nanocrystals (~300% increase from $B_T = 0$ to 0.13 T, Fig. 2c, 4), but it does not show a comparably strong field dependence in bulk CsPbBr$_3$ (~20% increase from $B_T = 0$ to 0.125 T).[8] This contrast may stem from confinement-enhanced hyperfine coupling in the nanocrystals.[24, 33, 48, 59-61] In this scenario, inhomogeneous hyperfine fields reduce $T_2^*$ more in nanocrystalline CsPbBr$_3$ than in bulk, and conversely, their suppression by magnetic fields has a greater impact on $T_2^*$ in the nanocrystals. We note that carrier spins within nanocrystals are also subject to hyperfine interactions with the surrounding nuclear spin bath, *e.g.*, from proton-bearing surface ligands, solvent, or polymer matrices,[62] distinguishing nanocrystals from bulk. Although $T_2^*$ is not necessarily related to optical coherence ($T_{2,o}$), it is intriguing that optical dephasing[16] and spin dephasing both occur with similar time constants of 50 – 80 ps in CsPbBr$_3$ nanocrystals at zero field. In other promising single-photon emitters like III-V nanocrystals, optical coherence is limited by nuclear-spin-flip dynamics.[63-65] If CsPbBr$_3$ nanocrystal optical coherence is similarly limited by spin-flip processes and carrier interactions with the nuclear spin bath, small magnetic fields may extend these optical coherence times, just as they extend the spin coherence times here (Fig. 2c, 4).

Finally, it is noteworthy that spin coherences are preserved at much higher temperatures in CsPbBr$_3$ nanocrystals than in bulk. Although the benefits of a magnetic field are restricted to temperatures below ~50 K, $T_2^*$ values of 80 ps at 100 K and 16 ps at room temperature are still observed in the nanocrystals. In comparison, $T_2^*$ in bulk CsPbBr$_3$ decreases precipitously above 20 K, dropping to ~20 ps at 100 K, above which it could not be measured.[8] CsPbBr$_3$ nanocrystals thus present greater opportunity for measurement and application at elevated temperatures. Furthermore, the different spin dephasing mechanisms active at various temperatures suggest potential strategies to engineer spin properties in metal-halide nanocrystals. For example, tuning or alloying the B-site cation[66, 67] to modify the hyperfine interactions may influence $T_2^*$ at low temperature or introduce new exchange contributions to the $g$ value. Similarly, modifying nanocrystal phonon energies through ion exchange, pressure, or shape control may allow extension of spin applications to elevated temperatures.

In summary, carrier spin-dephasing dynamics in metal-halide perovskite nanocrystals have been measured for the first time. Using TRFR spectroscopy, coherent spin precession of photogenerated holes in CsPbBr$_3$ nanocrystals has been observed from cryogenic temperatures up to room temperature. At low temperatures, spin coherence is limited by inhomogeneous hyperfine fields, but the spin-dephasing time can be extended dramatically by application of a small magnetic field. Under these conditions, $T_2^*$ approaches $\tau_{PL}$, and ~95% of emitted photons derive from excitons with coherent spin-polarized holes. Variable-temperature TRFR and PL measurements are consistent with additional spin dephasing at elevated temperatures driven by thermally activated LO phonons. The data highlight several major contrasts between spins in bulk and nanocrystalline $CsPbBr_3$. Overall, these results advance our fundamental understanding of spin dephasing in lead-halide perovskites and provide a basis for engineering such spin properties chemically or with magnetic fields, potentially opening the door to enticing spintronic and quantum information applications.

**Supporting Information**
The Supporting Information is available free of charge at http://pubs.acs.org and includes additional details about nanocrystal synthesis and variable-temperature PL lifetimes, TRFR traces used in Figure 2 and 4 as well as TRFR traces from replicate samples, and FFTs of TRFR traces (PDF).

The authors declare no competing financial interest.

**Acknowledgements**
This research was supported by the National Science Foundation (NSF) through DMR-1807394 and through the UW Molecular Engineering Materials Center, a Materials Research Science and Engineering Center (DMR-1719797). MJC acknowledges support from the Washington Research Foundation through a Postdoctoral Fellowship. LMJ and YH acknowledge support from the Data Intensive Research Enabling Clean Technology (DIRECT) NSF National Research Traineeship (DGE-1633216). Part of this work was conducted at the Molecular Analysis Facility, a National Nanotechnology Coordinated Infrastructure site at the University of Washington that is supported in part by the National Science Foundation (ECC-1542101), the University of Washington, the UW Molecular Engineering & Sciences Institute, and the UW Clean Energy Institute. The authors thank Prof. Devin Mackenzie for thoughtful input and support.

**TOC Graphic**

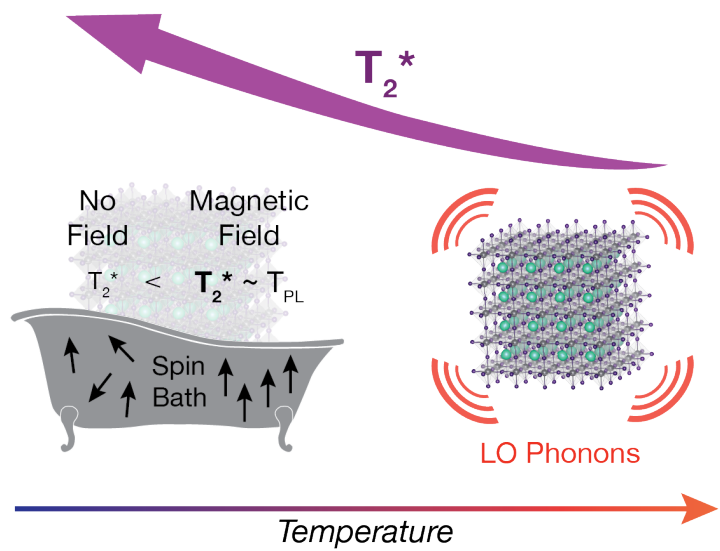

# Coherent Spin Precession and Lifetime-Limited Spin Dephasing in CsPbBr$_3$ Perovskite Nanocrystals

Matthew J. Crane,[†,‡] Laura M. Jacoby,[†,‡] Theodore A. Cohen,[‡,§,∥] Yunping Huang,[§]

Christine K. Luscombe,[‡,§,∥] Daniel R. Gamelin[‡,∥,]*

[‡]*Department of Chemistry, University of Washington, Seattle, WA 98195-1700*

[∥]*Molecular Engineering & Sciences Institute, University of Washington, Seattle, WA 98195-1652*

[§]*Department of Materials Science & Engineering, University of Washington, Seattle, WA 98195-2120*

*Email: gamelin@chem.washington.edu

[†]*These authors contributed equally to this work.*

**Methods**

*CsPbBr$_3$ Nanocrystal Synthesis.* CsPbBr$_3$ nanocrystals were synthesized by methods adapted from ref. 1, 2. Briefly 1-octadecene (90%, Sigma Aldrich), Cesium oleate (synthesized from 99.9% Cs$_2$CO$_3$, Alfa Aesar), lead oleate (synthesized from 99.9% Pb(OAc)$_2$, Baker Chemical), and oleylamine (70%, Sigma Aldrich) were loaded into a three-neck flask and dried at 120°C under vacuum for 1 hour. The solution temperature was raised to 180°C and a premade solution of bromotrimethylsilane (97%, Sigma Aldrich) and 1-octadecene was swiftly injected into the reaction flask. The solution was allowed to react for ~ 5 s before quenching with an ice bath. During this time, the solution turned a vibrant chartreuse.

*ZFP3 Synthesis.* ZFP3, a zwitterion-functionalized polymer, was prepared as detailed in ref. 3. 2-(Dimethylamino)ethyl acrylate (143 mg, 1 mmol) and 1,1,1,3,3,3-hexafluoroisopropyl methacrylate (2124 mg, 9 mmol) was added to a vial that was pre-dried overnight in a 120 °C oven. 2,2'-azobis(2-methylpropionitrile) (8 mg, 0.5 mmol) and *tert*-nonyl mercaptan (8 μL, 0.5 mmol) were added subsequently. The vial was then degassed and sealed. The reaction was heated at 60 °C for one day, and at 80 °C for another two days. After cooling down to room temperature, the precursor polymer was dissolved in THF and dispersed into water. The precipitate was then washed with methanol and dried under vacuum overnight. Subsequently, the precursor polymer (400 mg) was dissolved in THF (8 mL) and heated to 60 °C. After 30 min, an excess of 1,3-propanesultone (400 mg, 3.28 mmol) was added dropwise. Methanol (3 mL) was added after 1 hour and additional methanol (3 mL) was added after 24 hours. The reaction was further heated at 60 °C for another day before cooling down to room temperature. The solution was transferred into a dialysis tube (molecular weight cut-off = 1000 g/mol) and stirred in methanol (1 L) for 2 hours and ethyl acetate (1 L) for an additional 2 hours. After dialysis, the solution inside the dialysis tube was transferred to a round-bottom flask and concentrated under vacuum to provide ZFP3 as a white solid in 81% yield.

*General Considerations.* For all spectroscopic measurements, the $CsPbBr_3$ nanocrystals were concentrated and transferred into solutions of ZFP3 to form dilute mixtures, unless otherwise specified. Nanocrystal/ZFP3 films were then formed by dropcasting these mixed solutions onto c-plane sapphire substrates. The room-temperature PLQY of the $CsPbBr_3$ nanocrystals in solution was >90%, and in ZFP3 films it was ~30% under the measurement conditions used here. Control experiments were also performed with nanocrystals in analogous PMMA films (see below). X-ray diffraction of the $CsPbBr_3$ was collected with a Bruker D8 Discover diffractometer with a IμS 2-D detector. Transition electron microscope images were collected using a Technai G2 F20 Supertwin TEM operating at 200 kV.

*Time-Resolved Faraday Rotation.* A schematic of the TRFR optics is provided in the Supporting Information as Figure S1. Pump and probe pulses for TRFR measurements were generated by passing the 776 nm output of a mode-locked Ti:Sapphire laser (Coherent MIRA-HP, pumped by a frequency doubled $Nd^{3+}$:YAG laser) operating at 76 MHz into an optical parametric oscillator to generate ~100 fs 515-520 nm pulses.[4] These pulses were split using a 50:50 beamsplitter and one portion was directed into a manual delay line. After the delay stage, the pump beam was directed through a photoelastic modulator (Hinds PEM-90) operating at 50 kHz to produce left and right circularly polarized light and focused onto the sample using a 300 mm lens at a 15° angle. The transmitted pump beam was blocked by an iris placed immediately after the sample. A 300 mm lens focused the linearly polarized probe beam onto the sample at a 0° angle with the sample to overlap with the pump beam, such that only the probe pulse was transmitted through the iris. The pump and probe beam diameters were both ~100 μm. Pump and probe powers were controlled independently using reflective neutral density filters and tuned so that the probe power was kept at 25% of the pump power. Rotation of the transmitted probe polarization was measured by passing the probe pulse through a Wollaston prism and focusing its vertical and horizontal components onto a split silicon photodiode. A half-wave plate was inserted before the Wollaston prism to ensure that the photodiode was well balanced in the absence of the pump. The signal from the photodiode was measured using a lock-in amplifier (SR830) referenced to the 50 kHz photoelastic modulator. System artifacts were reduced by collecting background traces using only the probe pulses under otherwise identical conditions. The sample temperature and transverse magnetic field were controlled *via* a Janis optical cold-finger cryostat and an electromagnet. The accuracy of the thermometry at the sample position has been independently verified by magnetic circular dichroism spectroscopy on analogous simple paramagnets (showing 1/T intensity variation) mounted in the same configuration.[5] Signal was optimized by tuning the pump/probe wavelength before collecting data. The TRFR data did not change significantly when varying the pump power from 1.0 to 50 mW cm$^{-2}$.

The pump and probe pulse fluences used in these measurements (*e.g.*, $6 \cdot 10^{-3}$ and $2 \cdot 10^{-2}$ excitations per nanocrystal per pulse, respectively, for the data in Fig. 2) are orders of magnitude smaller than those reportedly required for trion formation in $CsPbBr_3$ nanocrystals at room temperature and using similar repetition rates.[6] The continuity of our TRFR data across all temperatures suggests that we are monitoring the same carriers at both low and high temperatures.

*Photoluminescence, absorption, and related measurements.* For time-resolved and variable-temperature photoluminescence measurements, the sample was excited using either the TRFR pump beam or a 375 nm Edinburgh diode laser with a repetition rate of 100 ns. Photoluminescence signals were measured using a streak camera mounted on a 0.1 m monochromator (150 g/mm blazed at 600 nm). Room-temperature CW photoluminescence data were collected using a 0.5 m monochromator (150 g/mm blazed at 500 nm) equipped with a CCD. Absorption spectra were measured using a Cary 5000 spectrophotometer. The sample PLQY was measured using an integrating sphere and multichannel analyzer (Hamamatsu, C9920-12).

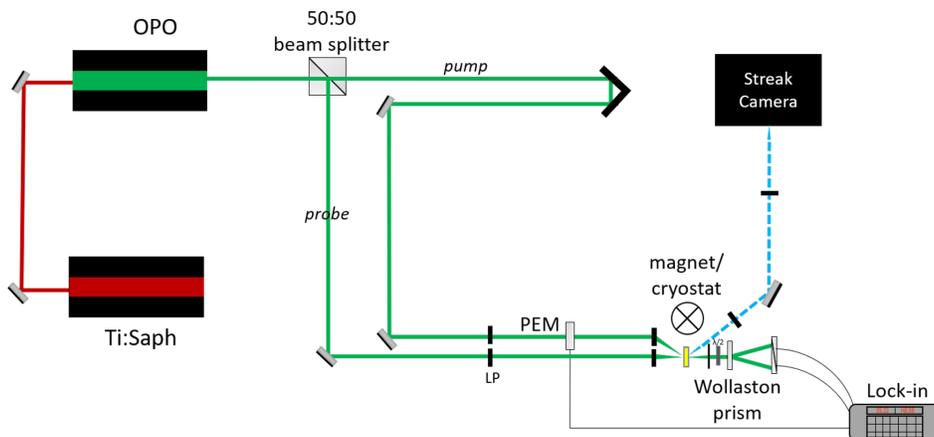

**Figure S1.** Schematic of the TRFR optical configuration used here, as described in the Methods section of the main text. A 76 MHz Ti:Sapphire generates a 3.8 W train of ~100 fs pulses, which are converted to ~520 nm light through an optical parametric oscillator. These visible pulses are split into a pump path, which traverses a mechanical delay line and is modulated between right and left circular polarizations using a PEM, and a probe path. Both pulses are focused onto the $CsPbBr_3$ nanocrystal/polymer film in a cryostat using 300 mm lenses to an ultimate stop size with a diameter of 100 μm. The cryostat is mounted between the poles of an electromagnet. The pump beam is blocked by an iris placed after the sample, and the probe beam is split by a Wollaston prism and the two components are detected by a balanced photodiode and recorded using a lock-in amplifier triggered by the PEM. The PL data reported in the main text were collected using exactly the same setup and conditions by directing emission from the sample into a monochromator with an attached streak camera.

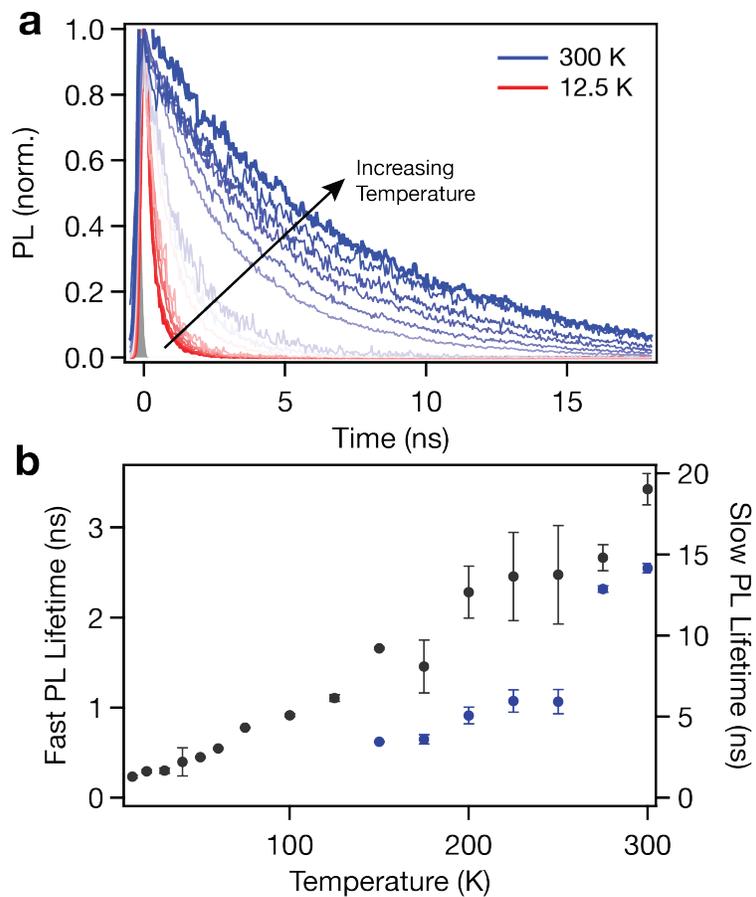

**Figure S2.** Variable-temperature CsPbBr$_3$ nanocrystal/polymer thin-film PL decay dynamics. **(a)** PL decay from CsPbBr$_3$ nanocrystal/ZFP3 polymer thin films, ranging from room temperature to 12.5 K, and **(b)** decay times obtained by fitting the data from panel (a). PL decay curves between 12.5 and 125 K were fit with monoexponential functions, and from 150 to 300 K with biexponential functions. The shaded peak at 0 ns in panel (a) shows the instrument response function. $\lambda_{ex}$ = 375 nm.

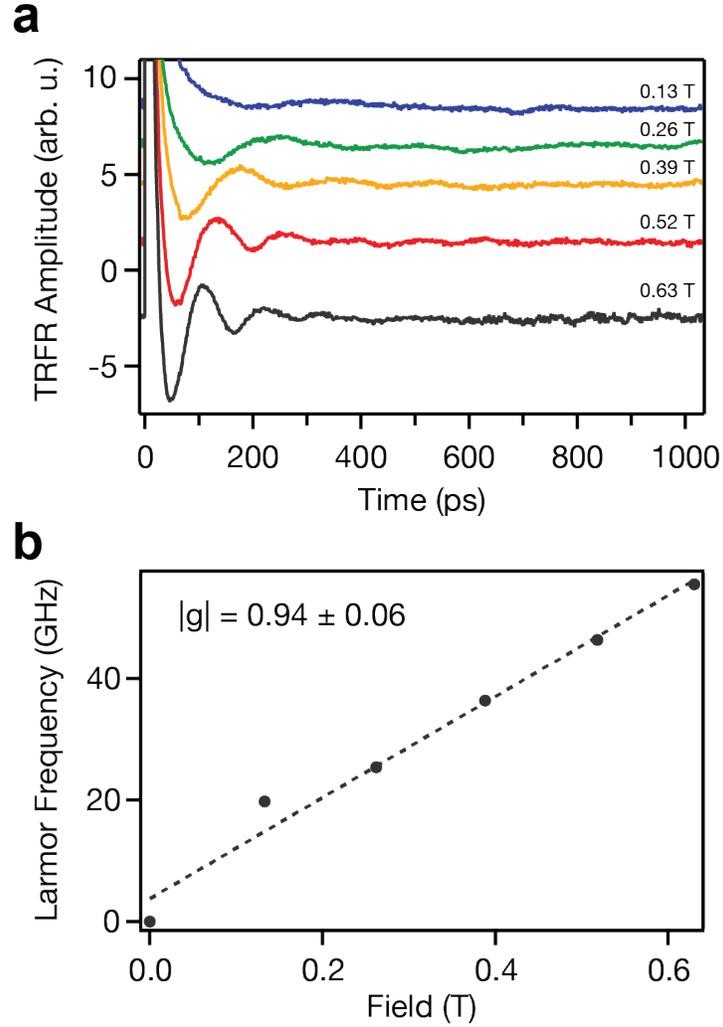

**Figure S3.** Spin dynamics of CsPbBr$_3$ nanocrystals in ZFP3 thin films at 20 K. **(a)** TRFR traces of CsPbBr$_3$ nanocrystals measured at different applied transverse magnetic fields, showing an oscillating signal with monotonically increasing Larmor frequency as a function of increasing magnetic field. **(b)** Plot of the Larmor frequencies *vs* magnetic field, obtained by fitting the TRFR traces in panel (a) using eq 1 of the main text. These data are linear and yield a |g| value of 0.94 ± 0.06, similar to the value of 0.88 measured for the same CsPbBr$_3$ nanocrystals in ZFP3 polymer and presented in the main text.

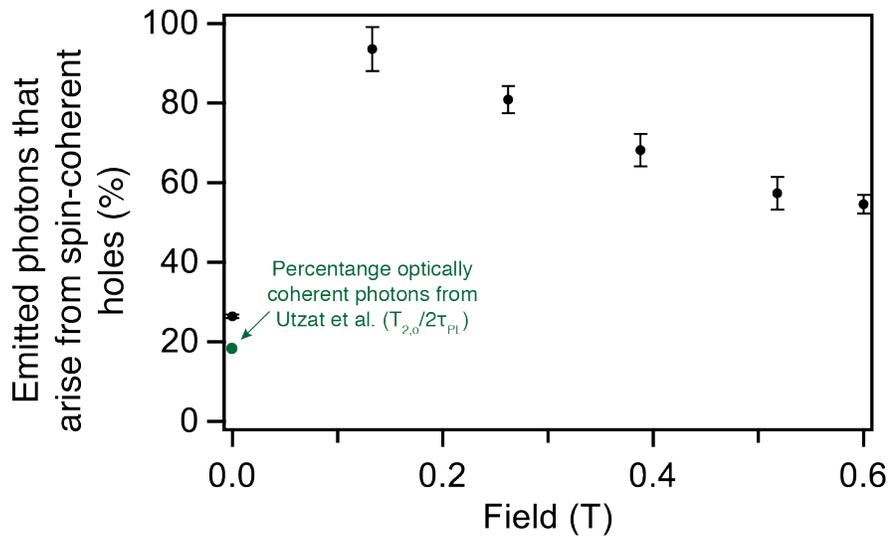

**Figure S4.** The percentage of emitted photons that arise from recombination of spin-coherent holes ($T_2^*/\tau_{PL}$), plotted *vs* applied magnetic field (black). For comparison, the green data point indicates the percentage of emitted photons coming from optically coherent excitons ($T_{2,o}^*/2\tau_{PL}$), taken from Utzat *et al.*[7]

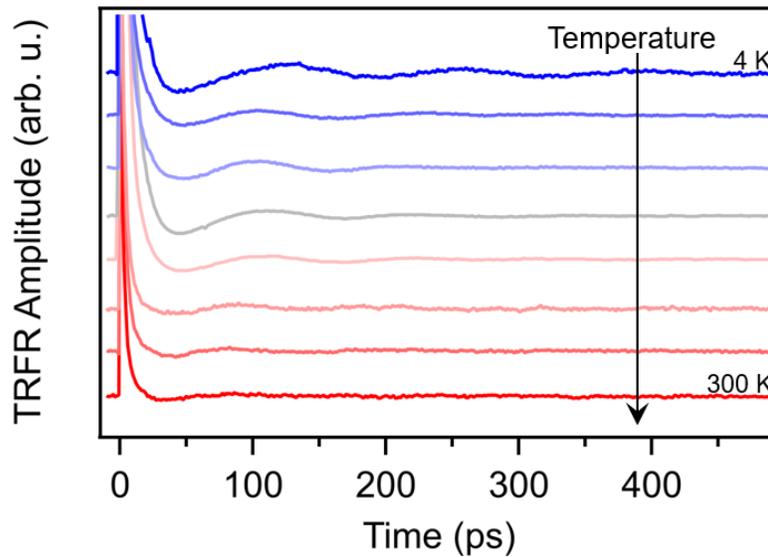

**Figure S5.** Variable-temperature TRFR traces at $B_T = 0.60$ T used for fits in Figure 4 of the main text.

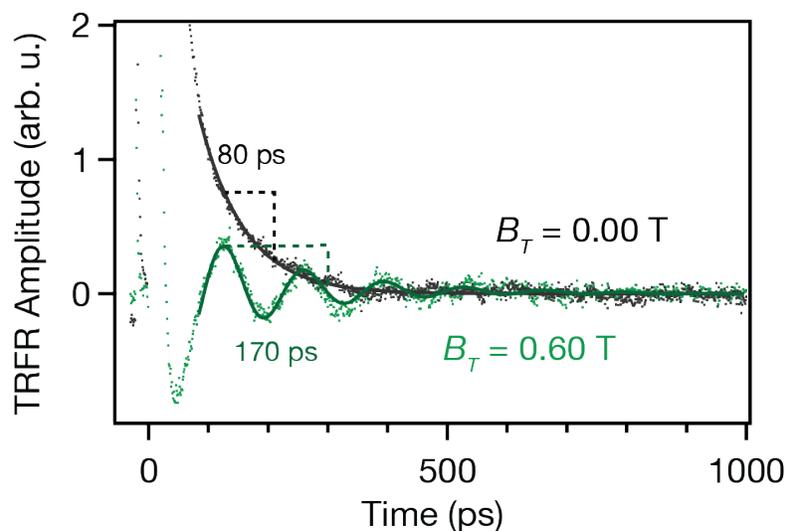

**Figure S6.** A comparison of TRFR traces collected with and without an applied magnetic field at T = 4.5 K from Figure 2a in the main text. Solid lines show fits to the data using eq. 1, while the dashed lines illustrate the time for the sample to decay to 1/e amplitude, 80 ps for $B_T = 0.00$ and 170 ps for $B_T = 0.60$. Under an applied magnetic field, the nuclear spin bath becomes polarized, suppressing inhomogeneous magnetic (hyperfine) fields that accelerate dephasing.

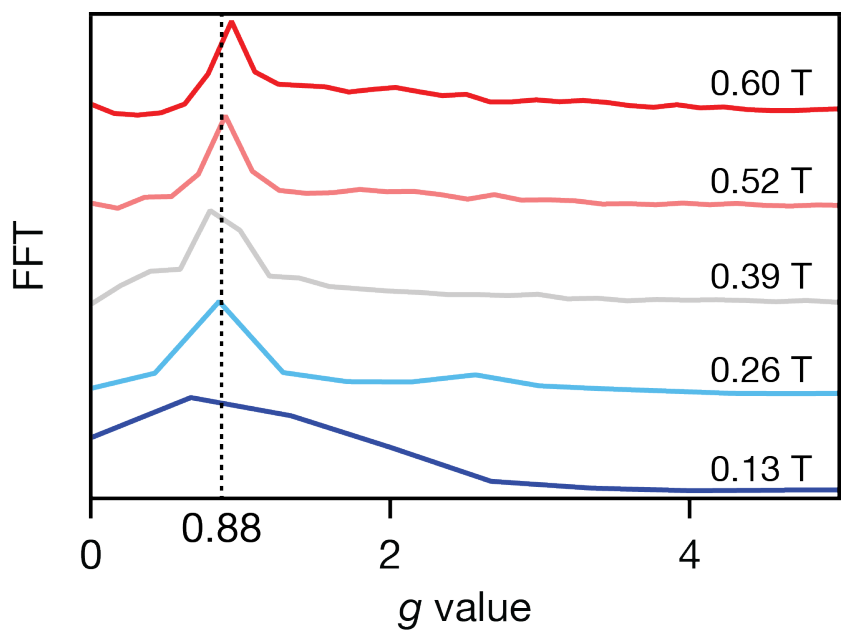

**Figure S7.** Fast Fourier transforms of TRFR traces (from Figure 2a in the main text) from CsPbBr$_3$ nanocrystals in ZFP3 at 4.5 K collected under various applied transverse magnetic fields, showing one prominent oscillating species. The dashed line illustrates the $g$ value fit from Figure 2c of $g = 0.88 \pm 0.01$. The FFT peak width increases as the applied magnetic field decreases because precession frequency decreases, lowering the number of full oscillations before the polarized spins decay.

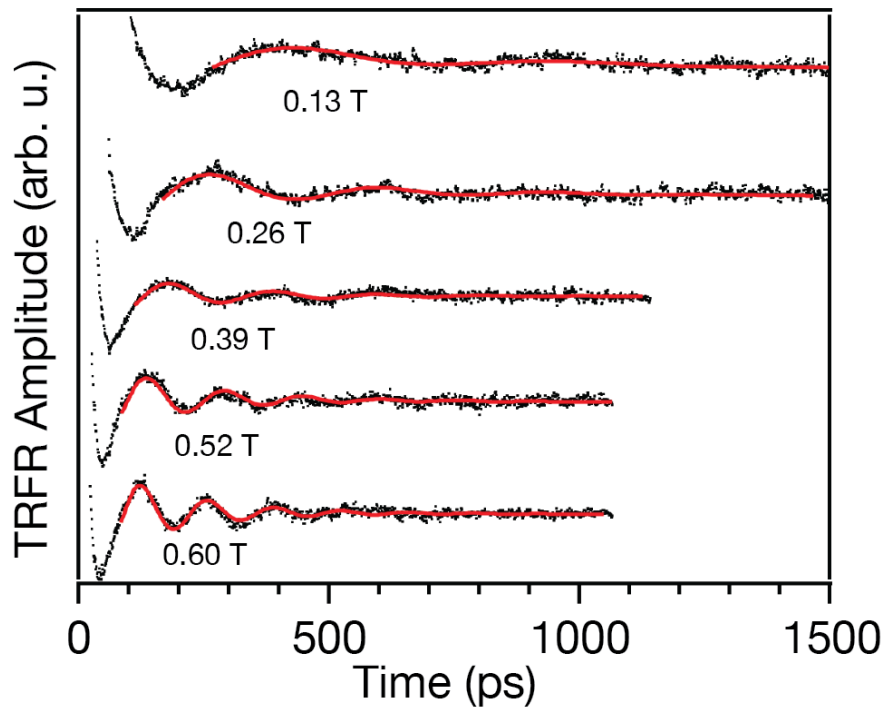

**Figure S8.** TRFR traces of CsPbBr$_3$ nanocrystals in a ZFP3 polymer matrix under various applied transverse magnetic fields, from Figure 2a of the main text. Fits of these data to eq. 1 of the main text are shown in red lines.

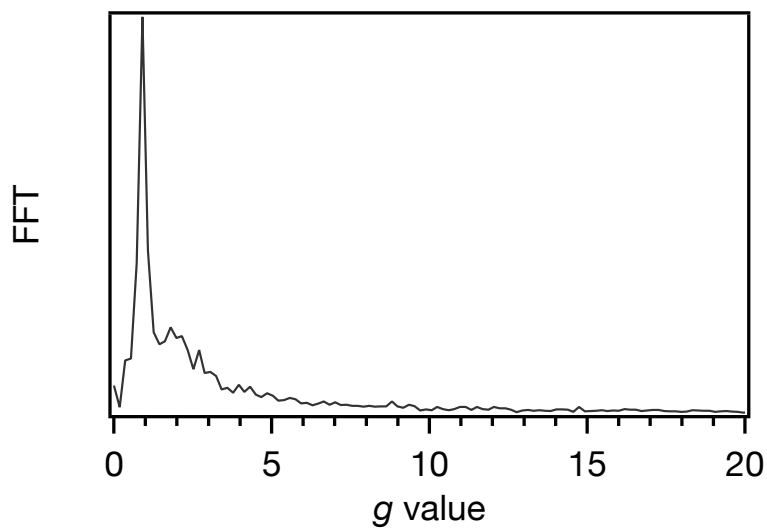

**Figure S9.** The $B_T$ = 0.52 T data from Fig. S8, replotted on a broader *x* scale. The shoulder at $|g| \sim 2$ is similar to the feature observed in ref. [8] and assigned to precessing electrons in bulk CsPbBr$_3$, but it is too weak here to be interpreted with confidence.

**Table S1.** Comparison of literature $g$ values for $CsPbBr_3$ (bulk and nanocrystal) with those reported here.

| | $g_{exciton}$ | $g_e$ | $g_h$ | $g_{trion}$ |
|---|---|---|---|---|
| Fu et al.[9] (individual NCs) | ~2.3 ± 0.4 (PL) | ~2.0 (assumed) | ~0.4 (assumed $g_e$ and eq S1) | ~2.4 ± 0.3 (PL) |
| Canneson et al.[10] (NC ensembles) | --- | 2.2 (from eq S1) | 0.22 (MCD+PL) | 2.4 (MCD) |
| Belykh et al.[8] (bulk) | 2.4 (MCD) 2.7 (from eq S1) | 1.96 (TRFR) | 0.75 (TRFR) | --- |
| Yu et al.[11] (effective mass calculations) | --- | ~2 | ~0.5 | --- |
| This report (NC ensembles) | ~2.6 (from eq S1) | ~1.8 (TRFR) | ~0.8 (TRFR) | --- |

These studies all assume the following relationship:

$$g_{exciton} = g_e + g_h \quad (S1)$$

Table S1 shows that our $g$ values and assignments are consistent with literature. For example, Fu et al.[9] report $g_{exciton}$ values from individual nanocrystals over a rather broad range, from ~2.0 to 2.7. By their analysis (which assumes eq S1), $g_h$ should correspondingly range from ~0.3 to ~0.7. Our $g_h$ is therefore similar to their $g_h$ but far from $g_{trion}$, for example. Similarly, for our data, eq S1 yields $g_{exciton}$ ~ 2.6, which is squarely within the range reported by Fu et al. Noting that the results of Belykh et al.[8] call into question the accuracy of eq. S1 (from MCD, they measured $g_{exciton}$ = 2.35, but from eq S1 they obtained $g_{exciton}$ = 2.7), Table S1 shows that our $g$ values are generally consistent with experimental literature. Additionally, our values are also consistent with predictions from effective-mass calculations ($g_e$ ~ 2.0, $g_h$ ~ 0.5).[11]

**Table S2.** Isotopic abundance and hyperfine coupling parameters for elements in $CsPbBr_3$.[8]

| Isotope | Abundance (%) | Nuclear Spin | Hyperfine Coupling Constant (µeV) |
|---|---|---|---|
| $^{133}Cs$ | 1 | 7/2 | - |
| $^{204}Pb$ | 1.4% | 0 | - |
| $^{206}Pb$ | 24.1% | 0 | - |
| $^{207}Pb$ | 22.1% | -1/2 | $A_h$ = 100 |
| $^{208}Pb$ | 52.4% | 0 | - |
| $^{79}Br$ | 50.7% | -3/2 | $A_e$ = 20 |
| $^{81}Br$ | 49.3% | -3/2 | $A_e$ = 20 |

The inhomogeneous spin-dephasing time for precession of free carriers in a nuclear hyperfine field at low temperatures can be expressed using Eq. S2:[12, 13]

$$T_2^* = \hbar \sqrt{\frac{3N_L}{16 \sum_j I^j (I^j + 1)(A^j)^2 y_j}} \quad \text{(S2)}$$

Here, $N_L$ is the number of nuclei within the localization volume of the precessing carrier. The sum runs over all types of nuclear isotopes in the lattice, $j$, with their respective nuclear spins, $I^j$, hyperfine constants, $A^j$, and natural abundances, $y_j$. It is informative to consider how the hyperfine dephasing in $CsPbBr_3$ compares to other relevant materials, such as CdSe and GaAs with identical numbers of nuclei in the relevant volume. Using Eq. S2 we estimate that hyperfine-driven dephasing is ~3 times faster in $CsPbBr_3$ than in CdSe. Similarly, hyperfine-driven dephasing is ~3 times faster in GaAs than in $CsPbBr_3$. For these calculations, we use the compiled elemental properties in ref. [13] and Table S2. We note that this analysis predicts a longer $T_2^*$ in $CsPbBr_3$ (~1 - 2 ns) than we observe experimentally, which may in part reflect population decay due to the short exciton lifetimes. These results highlight the impact of hyperfine-induced hole-spin dephasing in $CsPbBr_3$ attributable to lattice ionicity and strong Fermi contact via the Pb(6s) orbital character of the valence-band edge.

**Table S3.** Sample statistics for $g$ and $T_2^*$ values in different $CsPbBr_3$ nanocrystal films at 4.5 K.

| Compound | $g$ value | $T_2^*$ (0.13 T) |
|---|---|---|
| $CsPbBr_3$ in ZFP3, sample A | 0.88 ± 0.01 | 292 ± 17 ps |
| $CsPbBr_3$ in ZFP3, sample B | 0.94 ± 0.06 | 319 ± 20 ps |
| $CsPbBr_3$ in PMMA | 1.02 ± 0.02 | 314 ± 43 ps |

**Table S4**. A comparison of literature values for the acoustic phonon coupling constant, σ, in metal-halide perovskite nanocrystal ensembles derived from the independent Boson model (eq 4 in the main text).

| | Nanocrystal Samples | Acoustic coupling constant, σ (μeV/K) |
|---|---|---|
| Shinde et al. [14] | $CsPbBr_3$ | 23 |
| Saran et al. [15] | $CsPbBr_3$ | 33 |
| **This report** | $CsPbBr_3$ | **60 ± 20** |
| Lohar et al. [16] | $CsPbCl_3$ | 81.3 |
| Ghosh et al. [17] | $FAPbBr_3$ | 100 |
| Ai et al. [18] | $CsPbBr_3$ | 152 |

The uncertainty specified with our acoustic coupling constant reflects the fact that the fit depends very weakly on the acoustic coupling constant, as noted by Utzat et al.[7] It is noteworthy that single-nanocrystal measurements consistently show substantially smaller values of σ (<10 μeV/K)[19,20] than found in ensemble measurements. As highlighted in Ref. 14, σ depends on nanocrystal size, and ensemble size distributions may contribute to this difference.